| | |
|---|---|
| Title | **Promoting lentil germination and stem growth by plasma activated tap water, demineralized water and liquid fertilizer** |
| Authors | S. Zhang, A. Rousseau, T. Dufour |
| Affiliations | LPP, CNRS, UPMC Univ. Paris 06, Ecole Polytech., Univ. Paris-Sud, Observatoire de Paris, Université Paris-Saclay, Sorbonne Universités, PSL Research University, 4 Place Jussieu, 75252 Paris, France. E-mail: thierry.dufour@upmc.fr |
| Ref. | RSC Adv. ,2017, 7, 31244-31251 |
| DOI | 10.1039/c7ra04663d |
| Abstract | Tap water, demineralized water and liquid fertilizer have been activated using an atmospheric pressure plasma jet (APPJ) to investigate their benefits for the germination rate and stem elongation rate of lentils from Puy-en-Velay (France). By plasma-activating tap water, we have obtained germination rates as high as 80% (instead of 30% with tap water). Also, higher stem elongation rates and final stem lengths were obtained using activated tap water compared with commercial fertilizer. We show that these rates of germination and stem growth strongly depend on the combination of two radicals generated in the liquids by the plasma: hydrogen peroxide and nitrate. This synergy appears to be a condition for releasing seed dormancy through the endogenous production of NO radicals |

# Introduction

Cold atmospheric plasmas (CAP) have brought a wide variety of applications such as material processes [1, 2], biomedical applications [3-6], micro-fabrication [7]. They can be considered as cocktails containing chemical species and ions, broadband radiation, and even electric field making them effective and successful in such kinds of applications.

CAP applications in agriculture have recently emerged due to the predicament in modern agriculture. Intensive agriculture has the advantage of large yield at the expense of large input such as more employment of workers, larger nonorganic fertilizer and pesticide that lead to worse environment pollution and more pesticide residue. As for the organic agriculture, this is now still contentious from the perspective of yield [8]. Plasma agriculture, on the other hand, could be a solution to the current problem. As shown below, cold plasma could promote seeds germination and stimulate plants growth on some crops without nonorganic fertilizer. Additionally, it is also verified that the atmospheric pressure plasma jet can cure fungus of plants and improve plants resistance to some diseases, which means less use of pesticide [9]. In this respect plasma agriculture stands for a great potential to solve the problem of modern agriculture.

As for plasma treatment on seeds, nearly all the works are performed by directly treating seeds or even plants by CAP [10]. One advantage of direct treatment is that the short-lived species such as O, OH, NO, and even the electric field or UV produced by the plasma [11, 12] could be possibly effective; some reports indicate this type of plasma treatment could enhance or modify the germination and/or growth of the oilseed rape seeds [13], bean and wheat seeds [14–16]. Direct plasma treatments on tomato, radish and mulungu gave also positive effects [17-20]. On the other hand, plasma activated liquids (mainly different types of waters) provides a much greater degree of mobility of CAP used in agriculture and could be much more convenient in the industry. Even if few studies report so far plasma processes dedicated to the activation of liquids for agriculture applications, Khacef et al have recently developed a process to improve germination rates of radishes, tomatoes and sweet pepper [21-23]. Hence, they obtain an increase of the germination rate as high as 80% in the case of radishes while stems elongation of 60% is obtained for tomatoes.

In this work, tap water, demineralized water and liquid fertilizer are used to irrigate seeds and seedlings from germination to early stages growth. These liquids are also exposed to an atmospheric pressure plasma jet (APPJ) to obtain plasma activated liquids (PAL). If tap/demineralized waters have already shown differences on plants growth [24], we investigate here how these PAL can have an important impact on germination rate as well as on stems elongation. Two main radicals in the liquid phase are characterized: nitrates and hydrogen peroxide. Their respective effects on the previous agronomical parameters are discussed.

# Experimental setup

The plasma source utilized for liquids activation is an APPJ as shown in Figure 1. The APPJ is composed of a quartz dielectric tube (inner diameter = 4.5 mm, outer diameter = 8 mm) with an inner rod electrode (AC high voltage) and an annular grounded electrode. The AC high voltage is delivered by a function generator (ELC Annecy France, GF467AF) and is amplified by a power amplifier (Crest& Audio, 5500W, CC5500). A ballast resistor (250 kΩ, 600W) protects the power supply from too large currents that may result from plasma glow-to-arc transitions. A capacitor (10 nF) is connected to the grounded electrode to monitor the plasma current. The APPJ is supplied with 500 sccm helium controlled by a flow controller (MKS, model 1259C - 00500SV). A glass vessel is placed 5-10 mm downward the APPJ to collect liquids with typical volume of 10-20 mL. After its activation, this liquid is immediately poured onto the seeds or seedlings.

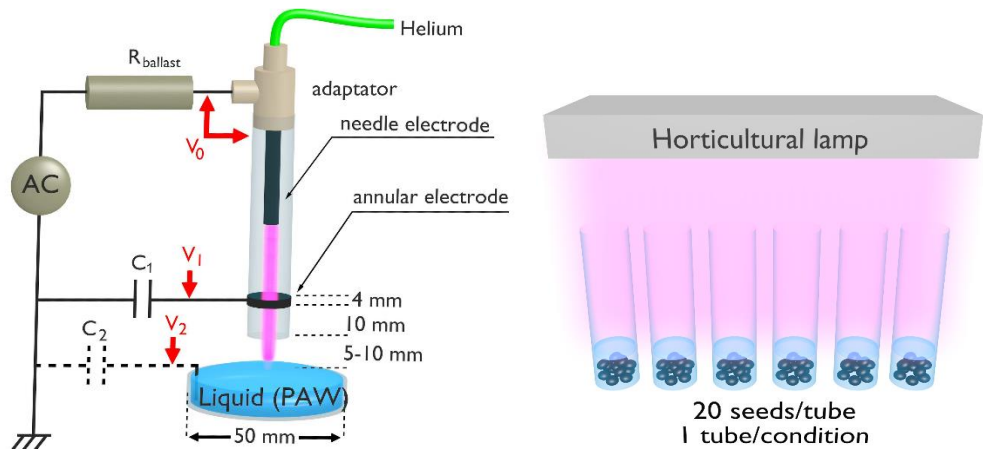

*Figure 1. (a) Schematic representation of the APPJ dedicated to the activation of liquids (b) Schematic representation of the artificial lighting for six test tubes, each one containing 20 seeds and allocated to a specific experimental condition*

As reported in Table 1, six experimental conditions are considered (three PALs and their respective Controls), each one corresponding to a test tube with 20 seeds (lentils from Puy-En-Velay, France). The chemical composition of tap water is reported in Table 2. It has been evaluated on 63 samples by the Agence Régionale de Santé (Ile-de-France region) on Public networks and interior building networks from the 5[th] district of Paris city, in February 2017 [25]. It is important to notice first the presence of nitrates (30.7 mg/L) and second that the charge in micro-organisms such as *Enterococcus* and *Escherichia Coli* can be considered as negligible. The liquid fertilizer is a universal product commercialized by Truffaut©, dedicated to indoor plants and utilized following the supplier recommendations (dilution to 0.5%). The fertilizer brings the essential elements (nitrogen, phosphorus, potash) to the soil to feed seedlings roots. The micro-organisms in the soil transform these elements into mineral salts, which can then be assimilated by the seedlings to constitute their nutritional intakes.

The seeds are daily irrigated with different liquid volumes depending on time: (i) at days 1 and 2, 10 mL/day is poured in each test tube (and removed the day after) so that all the seeds are immerged and (ii) after day 3, 1 mL/day of liquid is poured in each test tube without being removed the day after. The experiments are performed at room temperature (22.5°C±0.5°C) and the relative humidity ranges between 35 and 45%. Lighting is provided by the ambient light and reinforced 6 hours a day using a horticultural dimmable LED panel. The emission spectrum of this lamp is indicated in Figure 2; it presents two main peaks at 460 and 636 nm, a continuum in between and two minor peaks at 732 and 912 nm. Finally, it is important to underline that no soil or any organic substrate is utilized to clearly evidence the direct impact of PAL on seeds and seedlings, without any potential interference resulting from soils or solid carbon sources.

| TUBE NB. | ACRONYM | SIGNIFICATION |
|---|---|---|
| 1 | TAP Ctrl | Tap water as Control |
| 2 | TAP PAL | Tap water as Plasma Activated Liquid |
| 3 | DEM Ctrl | Demineralized water as Control |
| 4 | DEM PAL | Demineralized water as Plasma Activated Liquid |
| 5 | FTZ Ctrl | Liquid fertilizer as Control |
| 6 | FTZ PAL | Liquid fertilizer as Plasma Activated Liquid |

*Table 1. The six experimental conditions investigated*

| PARAMETER | UNIT | MIN | MEAN | MAX | REF. LIMIT OF QUALITY |
|---|---|---|---|---|---|
| pH | pH | 7.3 | 7.7 | 8.1 | 6.5-9 |
| Turbidity | NFU | 0.0 | 0.1 | 0.5 | 2 |
| Free chlorine | mg/LCl2 | 0.1 | 0.2 | 0.3 | - |
| Nitrates | mg/L | 23.6 | 30.7 | 37.4 | 50 |
| Conductivity | µS/cm | 520.0 | 571.9 | 637.0 | 200-1100 |
| Ammonium | mg/L | 0.0 | 0.0 | 0.0 | 0.1 |
| *Escherichia coli* | n/100 mL | 0.0 | 0.0 | 0.0 | 0.0 |
| *Enterococcus* | n/100 mL | 0.0 | 0.0 | 0.0 | 0.0 |
| Sulfato-reducing bacteria | n/100 mL | 0.0 | 0.0 | 0.0 | 0.0 |
| *Coliforma* bacteria | n/100 mL | 0.0 | 0.0 | 0.0 | 0.0 |

*Table 2. Biochemical analysis of tap water from public networks and interior building networks measured in the 5$^{th}$ district of Paris (Feb. 2017) by the Agence Régionale de Santé [25]*

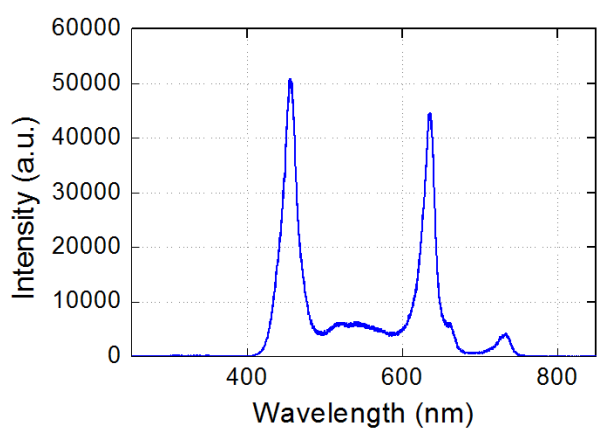

*Figure 2. Emission spectrum of the horticultural lamp*

Two long-life radicals have been quantified in the liquid phase: hydrogen peroxide species and nitrate ions. For the detection of $H_2O_2$, a solution of Titanium (IV) oxysulfate, 159.93 M, ~15 wt. % in dilute sulfuric acid, 99.99% (Sigma Aldrich) has been used. This acidic aqueous solution can then react in presence of $H_2O_2$, according to the following reaction

$$[Ti(OH)_3(H_2O)_3]^+_{(aq)} + H_2O_{2(aq)} \longleftrightarrow [Ti(O_2)(OH) \times (H_2O)_3]^+_{(aq)} + 2H_2O$$

Hence a yellow peroxotitanium complex $[Ti(O_2)OH(H_2O)_3]^+_{aq}$ is formed, with an absorbance peak measured at 409 nm [26]. Then the hydrogen peroxide concentrations can be calculated using the Beer's law. For the detection of $NO_3^-$, a PASCO scientific Nitrate Ion Selective Electrode has been utilized. It allows nitrate concentrations measurements on the $7.10^{-6}$ M-1M range for pH=2.5-11 and for temperatures between 273 K and 313 K. These concentrations measurements have not been performed *in situ* but *post mortem*, i.e. immediately after the plasma activation of the liquids. Some anions (e.g. $Cl^-$, $Br^-$, $ClO_3^-$) could interfere and cause electrode malfunction or drift, hence inducing an estimation error capped at 10%.

The impact of PAL on seeds and seedlings can be evaluated considering two distinct methodological approaches: a single step method where the growth rate ($R_{GROWTH}$) is calculated considering equation 1 and a two steps method where germination rate ($R_{GERM}$) and average stem elongation rate ($R_{STEM}$) are estimated according equations 2 and 3 respectively. In agriculture, $R_{GERM}$ describes how many seeds of a plant species, variety or seed lot are likely to germinate over a given period. This parameter does not depend on the size of the seedling but only on its appearance.

These two methods are commonly utilized in agronomy. The 1-step approach synthesizes the information through a single parameter while the 2-steps approach clearly deciphers germination phenomenon from "real" seedlings stems increase (see Table 3).

| METHODOLOGICAL APPROACH | RATES | EQUATIONS | | |
|---|---|---|---|---|
| Single step | Growth | $R_{GROWTH} = \dfrac{\sum \text{stems lengths}}{\text{Total number of seeds}}\bigg|_{tube} = \dfrac{\sum \text{stems lengths}}{20}$ | | [1] |
| Two steps | Germination | $R_{GERM} = \dfrac{\text{Number of seedlings}}{\text{Total number of seeds}}\bigg|_{tube} = \dfrac{\text{Final number of seedlings}}{20}$ | | [2] |
| | Stem elongation | $R_{STEM} = \dfrac{\sum \text{stems lengths}}{\text{Final number of seedlings}}\bigg|_{tube} = \dfrac{R_{GROWTH}}{R_{GERM}}$ | | [3] |

*Table 3. Methodological approaches and agronomical parameters*

# Results & Discussion

The results dealing with agronomy are presented, followed by measurements of reactive species in the liquid phases. Finally, we discuss how the measured agronomical parameters can be bridged with plasma chemistry.

## Agronomical parameters measurements

Photographs of lentils seeds and seedlings have been taken every day, along the entire experiment. As an illustration, the Figure 3 shows seedlings at days 6, 12 and 18 considering the six aforementioned conditions.

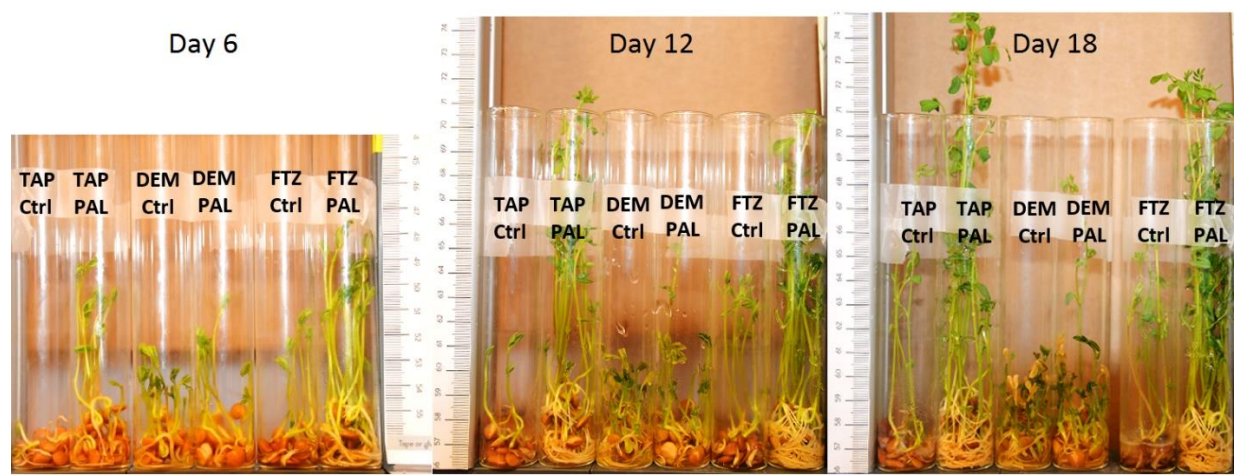

*Figure 3. Pictures showing lentils seeds and seedlings at day 6, 12 and 18. TAP: tap water, DEM: demineralized water, FTZ: liquid fertilizer, Ctrl: Control, PAL: plasma-activated liquid*

**[Single step methodological approach]**

PAL effect is evaluated through the growth rate (see equation 1), which corresponds to the ratio of seedlings stems sum on the total number of seeds contained in the test tube. This parameter is plotted in Figure 4 versus time for the six experimental conditions. Several observations and comparisons are noteworthy:

(i) When the seeds are irrigated with tap water (black curve, square symbols), seedlings grow quasi-linearly versus time to reach a mean value of 20 mm at day 15.

(ii) If seeds are now irrigated using the commercial fertilizer (cyan curve, star symbols), one can clearly observe a growth rate higher than the one obtained with tap water. This higher $R_{GROWTH}$ is particularly visible on the D6-D10 range. Beyond, $R_{GROWTH}$ tends to be limited and reach the same value as in the tap water case, i.e. 20 mm. The fertilizer has therefore a boosting effect at the early stages of the seedlings growth but not further.

(iii) The plasma-activated tap water treatment (red circles) is of major interest. $R_{GROWTH}$ drastically increases at the early seedlings stages (D6-D8 range) and reach a value as high as 90 mm at day 15, i.e. a value at least four times higher than when using tap water or even liquid fertilizer.

(iv) The plasma-activation of the liquid fertilizer (purple diamonds) induces a drastic increase of $R_{GROWTH}$ at early stages while this rise tends to slow down versus time and be limited to 90 mm, i.e. a value close to the one obtained using the plasma-activated tap water.

(v) With demineralized water – as control or plasma-activated – a growth rate close to 25 mm is reached at day 15. The absence of dissolute minerals seems not a hindrance for $R_{GROWTH}$ even if this rate is lower than those obtained using tap water and liquid fertilizer. Minerals could play an intermediate but crucial role in the interaction between seedlings roots and the plasma-induced radicals species in the liquid phase

(vi) In Figure 4a, the area comprised between fertilizer curve (cyan stars) and tap water curve (black squares) corresponds to the FTZ–TAP subtraction and stands for the gross contribution of the active elements contained in the liquid fertilizer, i.e. nitrogen species, phosphorus and potash. The benefit of mixing these active elements with plasma-activated tap water can be roughly estimated by adding the red circles curve to this difference, accordingly with the FTZ – TAP + PAL(TAP) curve plotted in Figure 4b (open red circles). For the sake of comparison, we have also reported in the same graph the curve of the plasma-activated liquid fertilizer. The two curves seem to fit, at least after 10 days, hence allowing us to claim: PAL(FTZ) ≈ FTZ – PAL + PAL(TAP). Therefore, if tap water initially enriched in nitrogen species, phosphorus and potash is then activated by plasma, it turns out that similar $R_{GROWTH}$ are obtained as for the fertilizer plasma activation. In Figure 4b, the overlapping of the two curves is less convincing between days 6 and 10. The resulting offset may be due to measurements uncertainties but could also indicate an unexpected effect in the liquid phase: an interaction between the plasma-induced radicals with the aforementioned active elements.

To summarize, if one compare the benefits of plasma activated tap water and liquid fertilizer, it turns out that the liquid fertilizer allows to boost $R_{GROWTH}$ only at early stages without offering the ability to increase final stem lengths while the plasma activation process allows to boost $R_{GROWTH}$ at early and late stages but also to increase final stem length. Finally, the plasma-activation of the liquid fertilizer unifies the two aspects: an early stage boost (probably due to the fertilizer) and an enhancement of $R_{GROWTH}$ (probably due to the plasma activated liquid).

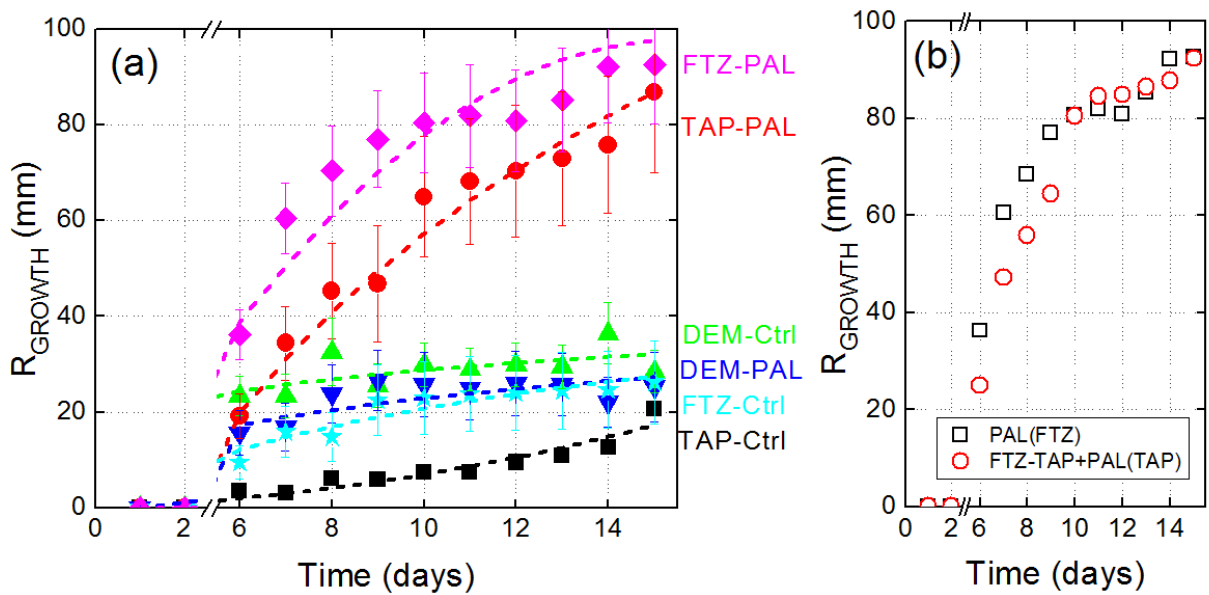

*Figure 4. (a) Growth rate parameter versus time. Seedlings are daily irrigated (b)$R_{GROWTH}$ comparison between plasma-activated fertilizer with a virtual solution corresponding to the operation FTZ – TAP + PAL(TAP)*

**[Two steps methodological approach]**

The germination rate is the ratio of seedlings (i.e. germinated seeds, whatever stems lengths) on the total amount of seeds contained in a test tube (see equation 2) and estimated at day 14. $R_{GERM}$ has been estimated in the six previous experimental conditions. Tap water and liquid fertilizer show similar trends: germination rate is lower than 40% in both cases while roofing close to 80% if the same liquids are plasma-activated. Surprisingly, this trend is not observed for demineralized water where surprisingly its plasma-activation tends to decay its germination rate from 71% to 54%.

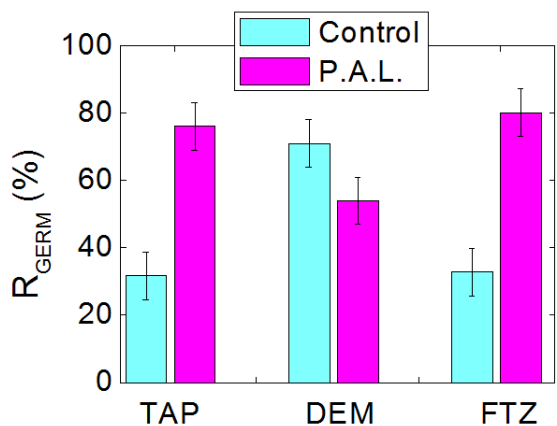

Figure 5. Germination rate of lentils seeds considering tap water, demineralized water and liquid fertilizer as Control or plasma activated media (day 14).

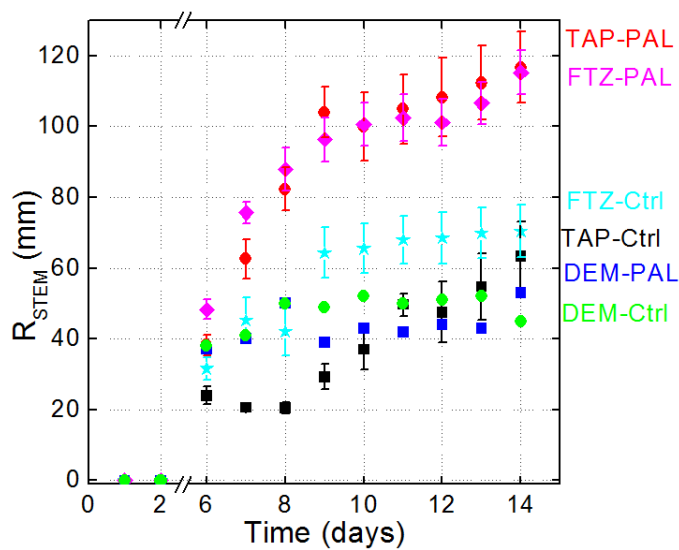

Figure 6. Stem elongation rate versus time of lentils seedlings considering the six different experimental conditions.

The stem elongation rate parameter is represented versus time in Figure 6. The observations reported on Figure 5 still apply here even if the trends are less pronounced. Also, and overall, the values of $R_{STEM}$ are slightly higher than those of $R_{GROWTH}$ since the number of considered seeds can be lower than 20.

## Chemistry of plasma activated liquids

Since seeds and seedlings have not been directly exposed to the plasma, only long lived chemical species contained in PAL can be involved in the enhanced rates of growth, germination and stem elongation. Since an exhaustive study of all the long-life species contained in the PAL is beyond the scope of the present article, we have focused our study on two of them often considered as important and distinct markers: hydrogen peroxide ($H_2O_2$) and nitrates ($NO_3^-$).

As shown in Figure 7a, no hydrogen peroxide is detected in the controls while an increase up to 160 µM is reached when tap and demineralized waters are plasma-activated. The absence of significant difference in $H_2O_2$ concentrations between these two media suggest that the presence of minerals (and eventually micro-organisms) in the tap water does not influence the production mechanisms of $H_2O_2$. In the case of the activated liquid fertilizer, the $H_2O_2$ concentration rises to 350 µM. Contrarily to tap and demineralized waters which are natively transparent, this liquid fertilizer presents a translucent appearance. Therefore, a dedicated calibration curve was performed so as to allow the utilization of spectrophotometry. As shown in Figure 7b, the nitrates can be detected in the 6 liquids. Two remarks are relevant:

(i) Nitrates concentration in the liquid fertilizer is very elevated: more than 10000 µM versus only 470 µM for the tap water and 50 µM in the demineralized water. In the case of control tap water, $[NO_3^-]=470$ µM which is a value close to the one officially communicated by the Paris Town Hall, namely 29 mg/L=467 µM [25].

(ii) The plasma exposure always increases nitrates concentrations in liquid phase but to different extents. Indeed, in tap water and liquid fertilizer, the relative enrichment in nitrate is 18% and 14% respectively while it is 286% in the demineralized water. This latter liquid is initially characterized by a very low electrical conductivity, which – according to figure 1 – behaves as a floating counter-electrode (instead of a grounded electrode in the two other cases. For this reason, overall radical production rates decrease in the plasma phase, inducing a lower absolute rise in nitrates when treating demineralized water.

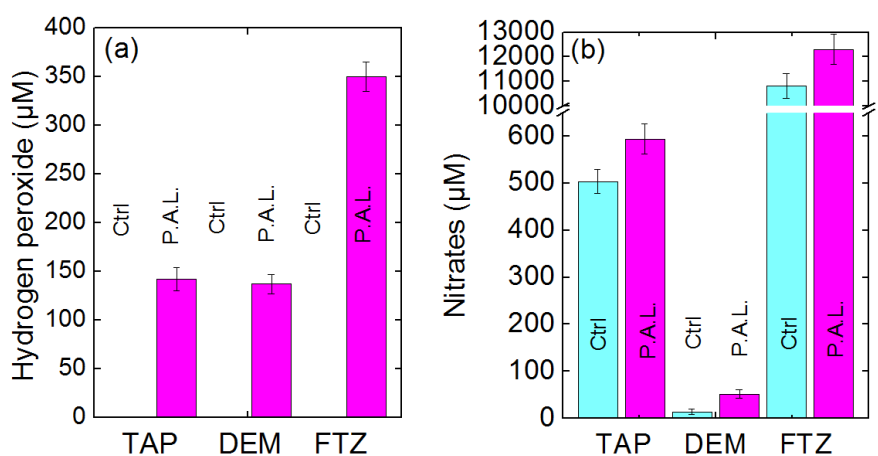

*Figure 7. (a) $H_2O_2$ and (b) $NO_3^-$ concentrations in Controls and PAL considering tap water, demineralized water and liquid fertilizer.*

## Bridging agronomical rates with plasma activated liquids chemistry

Dormancy and germination are jointly controlled by two main hormones: abscisic acid (ABA) and gibberellic acid (GA). ABA is a positive regulator of dormancy induction and a negative regulator of germination. GA releases dormancy, promotes germination and counteracts ABA effects [27]. The activity of these hormones can depend on exogeneous environment, in particular reactive species such as hydrogen peroxide, nitrates and nitric oxide.

Hydrogen peroxide is known to play a dual role in seeds physiological processes as well as in (a)biotic stress. As a signal molecule, it can promote the regulation of ABA and GA metabolism as well as hormonal balance [28]. According to the model of Liu reported in Figure 8, $H_2O_2$ can interrupt seeds dormancy through two pathways: the first corresponds to the enhancement of ABA catabolism and GA biosynthesis [29]. The signaling molecule (NO) does not regulate GA biosynthesis directly but instead acts as a temporary signaling molecule involved in the $H_2O_2$ regulation of ABA catabolism. Therefore, $H_2O_2$ mediates the up-regulation of ABA catabolism through NO signalization and GA biosynthesis. The second pathway indicates that a high production of ABA can inhibit GA biosynthesis. $H_2O_2$ could interrupt dormancy through GA signaling activation rather than influencing ABA metabolism [30].

Nitrate plays two main roles: (i) as a nutrient it is assimilated by plants enzymes (e.g. nitrate reductase) leading to the production of aminoacids and nitrogen compounds and (ii) as a signal molecule, it can control numerous aspects of plant development and metabolism [31]. The dormancy of viable seeds, i.e. germination failure although optimal environmental conditions, can sometimes be broken by imbibing seeds with nitrate solutions such as $NaNO_3$ or $KNO_3$ [32] [33]. These solutions promote the dormancy release of seeds only if $NO_3^-$ can mediate nitric oxide radicals. Sodium nitro-prusside (SNP) as well as cyanide (CN) and nitrite can also be considered as NO donors. At the opposite, the admixture of a NO scavenger such as c-PTIO extend seed dormancy [34]. In that respect, NO has to be considered as the key signaling molecule involved in seed dormancy loss, even if others may also participate such as hydroxylamine and nitrite. Also, it is worth mentioning that these radicals cannot be produced by cold plasma in the liquid phase, owing to their too short solubility [35]. They are biologically generated (e.g. with nitrate reductase) in the imbibed seed, i.e. at the interface between the seed and the (plasma activated) water.

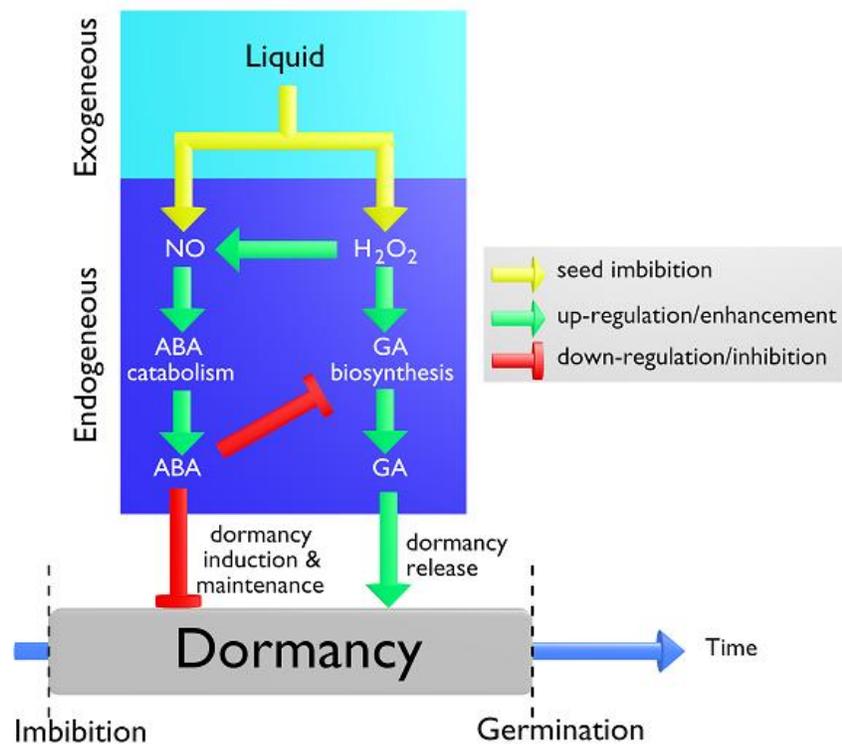

*Figure 8. Model of Liu showing how $H_2O_2$, NO, ABA and GA can regulate seed dormancy and germination [29]*

To bridge our agronomy results (germination and stem elongation rates) with P.A.L chemistry (measurements of $H_2O_2$ and $NO_3^-$), our discussion is focused on the comparison between tap and demineralized waters. In Figure 9, four data points corresponding to the four control/activated tap and demineralized waters are placed on X and Y axes considering their respective concentrations in hydrogen peroxide and nitrate. Horizontal purple arrows represent plasma activation process while vertical dashed orange arrows represent virtual pollution (mineralization and micro-organisms charging process). $R_{GERM}$ and $R_{STEM}$ global trends are reported to each of these four arrows.

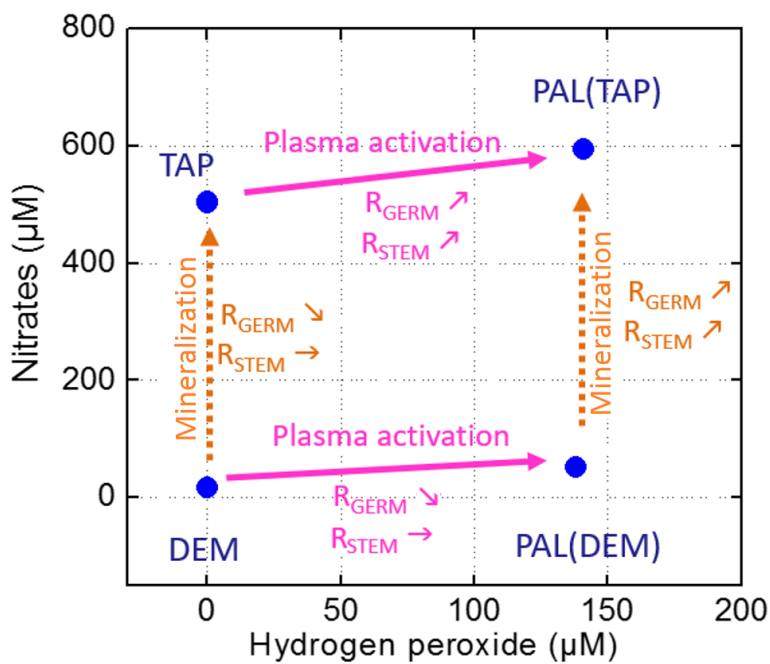

Figure 9. Synthesis graph bridging agronomical rates of lentils ($R_{GERM}$ and $R_{STEM}$) with plasma activation or virtual pollution processes. Control/PAL tap/demineralized waters are placed considering their concentrations in $H_2O_2$ and $NO_3^-$.

The plasma activation of demineralized water (bottom horizontal arrow) allows a strong increase in its $H_2O_2$ concentration (from 0 to 140 µM) while the increase in nitrate remains negligible comparatively. $H_2O_2$ seems inconsequential on stems elongation promotion since $R_{STEM}$ remains unchanged. However, the increase in $H_2O_2$ significantly lowers $R_{GERM}$. Considering the model of Liu, the increasing production of $H_2O_2$ in PAL could induce a disequilibrium between ABA and GA productions. Here, the seed imbibition only produces exogenous $H_2O_2$ radicals and no nitric oxide. Therefore, even if high concentrations of exogenous $H_2O_2$ radicals are produced they seem not involved into the production of endogenous radicals. As a result, the ABA catabolism remains low and ABA hormones at elevated levels so that dormancy is maintained. This first pathway is in competition with the second pathway where exogenous $H_2O_2$ radicals participate to the biosynthesis of GA to release dormancy. Finally, it turns out that ABA is stronger than GA, meaning that the first pathway prevails on the second, hence explaining why $H_2O_2$ the overall dormancy is maintained and incudes a lower germination rate [29].

The mineralization of deionized water (left vertical arrow) corresponds to an increase in nitrates (from 30 to 470 µM) while hydrogen peroxide remains equal to zero. This operation is accompanied by an unchanged $R_{STEM}$ parameter while $R_{GERM}$ is reduced, meaning that the germination process is promoted by a lower concentration in nitrates. At first sight, this inference seems in contradiction with conventional approaches where nitrate solutions are known to promote germination. However, one must bear in mind that these solutions do not result from water plasma activation but from conventional biochemistry methods, i.e. using ionic salts such as $KNO_3$ before being solubilized [33]. Moreover, and as previously stated, nitrate solutions shorten seeds dormancy only if $NO_3^-$ can mediate nitric oxide radicals. The increasing uptake in nitrates could enhance the production of endogenous NO radicals, inducing a higher activity of the ABA catabolism and therefore a lower amount of ABA hormones. This pathway could hence no more maintain dormancy. However, since demineralized as tap waters do not contain any $H_2O_2$ radicals, the simultaneous biosynthesis of GA is no more stimulated. According to our results, this second pathway prevails on the first pathway since dormancy is longer and germination rate lower. On the opposite, the obtained $R_{GERM}$ decrease may result from tap water chemical composition which has to be considered as a complex "cocktail" where radicals and minerals (e.g. ammonium, free chlorine, calcium, bicarbonates) induce a non-optimal ABA-GA balance, giving rise to a longest dormancy and therefore a delayed germination. Concerning the $R_{STEM}$ parameter which remains constant along the mineralization process (left vertical arrow), it turns out that seedlings elongations remain insensitive to radicals/minerals usually contained in tap water. The stem elongation mostly relies on the nutrients contained in their seeds.

The plasma activation of tap water (upper horizontal arrow) can be considered as an efficient process to produce $H_2O_2$ species but not for nitrates since their concentrations increase only from 550 µM to 600 µM, as shown in Figure 9. However, contrarily to the two previous processes, both $R_{GERM}$ and $R_{STEM}$ increase. If one refers to the Liu model in Figure 8, this dormancy release could result from low ABA and high GA concentrations. This stable disequilibrium could result from a strong up-regulation of the ABA catabolism mediated by $H_2O_2$ through a strong NO signalization, i.e. through an elevated production of endogenous NO radicals. This point is underlined by Rajjou et al., who clearly explain that nitrates do not affect seed dormancy on their own but rather act through NO biosynthesis [36]. These elevated amounts of biosynthesized nitric oxide would hence require a plasma activated liquid enriched both in $NO_3^-$ and $H_2O_2$ radicals. This 'two radicals' synergy appears as a mandatory condition even if it must be considered in a more complex chemistry where other long-life nitrogen radicals such as nitrites or peroxynitrites could also be involved at the interface of seeds and PAL.

## Practical aspects

In these research works, an APPJ device has been utilized to promote the germination and growth of lentils seeds by plasma-activating their irrigation liquids. Since a plasma jet requires a gas supply in argon or helium, an electrical source of energy (here AC high voltage) and since plasma jet can only activate small volumes of liquids, one may wonder if cold plasma technology is an approach intended to fail if one wants to transfer it to farms. Here, it is important to bear in mind that the finality of this work is to perform a proof of concept: demonstrating that plasma activation of water (and even of liquid fertilizer) can drastically boost germination and growth of an agronomic model: lentils. However, there are many technological locks that can be lifted to meet economic stakes: first, cold plasmas would rather have to be generated in dielectric barrier discharges (DBD) rather than in plasma jets. Indeed, no plasmagen gas is necessary in a DBD: the air could be used to directly generate almost the same reactive species in the liquid phase. Optimization work would nevertheless be necessary to process a larger surface of liquid, defining the optimal electrode-liquid interface gap and using dielectric barriers thinner and with lower dielectric permittivities. The resulting lower electrical energy consumption combined with green electricity coupling let us foresee a bright future for plasma agriculture.

Moreover, the plasma activation of the seeds and seedlings has been performed here following an indirect approach: water is first plasma-activated and second utilized to irrigate the seeds. Hence, only the long lifetime radicals such as nitrates and hydrogen peroxide can interplay with the physiologic mechanisms of the seeds. A direct approach could also be explored, consisting into immerging the seeds in water while performing their *in situ* activation. Thereby short lifetime radicals such as NO and OH could also participate into seeds germination.

# Conclusion

In this article, a proof of concept has been demonstrated where three plasma-activated liquids have shown distinct effects on same lentils seeds model.

Concerning the germination process, higher $R_{GERM}$ have always been obtained in the solely case of plasma activated tap water were both nitrates and hydrogen peroxide are present to concentrations as high as 500 µM and 150 µM respectively. This synergetic combination seems a prerequisite to induce the endogenous production of nitric oxide at the PAL-seed interface. According to the Liu mechanism, the resulting higher activity of the ABA catabolism would therefore reduce the down regulation on GA biosynthesis, hence releasing dormancy. The case of demineralized water activation is less obvious than the two others owing to its lower electrical conductivity which may change the plasma properties upon the activation time. Overall it appears that the solely production of nitrates or hydrogen peroxide using plasma activation process is not appropriate to enhance the germination of lentils seeds.

Concerning the stem elongation process, plasma activated tap water presents at least two main benefits compared to the utilization of a commercial liquid fertilizer: $R_{STEM}$ and final stem length (day 15) are both higher. We have shown that $H_2O_2$ alone cannot promote stems elongation. The best conditions to increase $R_{STEM}$ consists into plasma-generating nitrate and hydrogen peroxide in presence of minerals. This synergy appears as a mandatory condition even if not necessarily sufficient: other long-life nitrogen radicals such as nitrites or peroxynitrites might be involved. Furthermore, minerals could play a crucial role in this synergy, more specifically for nitrates production (see Figure 7) even if not for $H_2O_2$. Higher $R_{STEM}$ and final stem length PAL-induced are of interest and highlight the strong potential of PAL on the agronomical and economical points of views. Moreover, the species plasma-generated in the liquid phase could auto-degrade after a few hours while the liquid fertilizer can cumulate in the soil and induce long-term harmful pollution.

# Acknowledgements

This work has been achieved within the LABEX Plas@Par project, and received financial state aid managed by the Agence Nationale de la Recherche, as part of the Programme Investissements d'Avenir (PIA) under the reference ANR-11-IDEX-0004-02. Also, this work was supported by grants from Région Ile-de-France (Sesame, Ref. 16016309).